\documentclass[]{svjour3}
\usepackage{enumitem}
\usepackage{makecell}
\usepackage{lipsum}
\usepackage{url}
\usepackage{graphicx}
\usepackage{tikz}
 \usepackage{pgfplots}
\usepackage{multirow}
\usetikzlibrary{arrows,positioning,automata}
\usepackage{balance}
\usepackage{amsmath}
\usepackage[T1]{fontenc}
\usepackage{algorithm}
\usepackage{algpseudocode}
\usepackage{algpseudocode}
\usepackage{algorithm}
\usepackage{subfigure}
\usepackage{graphicx}
\usepackage[utf8]{inputenc}
\usepackage{balance}
\usepackage{color}

\usepackage{caption}

\usepackage{comment}
\usepackage{hyperref}
\usepackage{array}
\usepackage{soul}
\usepackage{parskip}
\newcommand{\eat}[1]{}

\begin{document}\sloppy

\title{Toward Maximizing the Visibility of Content in Social Media Brand Pages: A Temporal Analysis}
\titlerunning{Maximizing the Visibility of Content in Social Media Brand Pages}
\date{}
\author{Nagendra Kumar \and Gopi Ande\and J. Shirish Kumar\and Manish Singh\vspace{-2ex}}
\institute{\textit{Indian Institute of Technology Hyderabad, India – 502285}\\
\textit{Email: \{cs14resch11005, cs12b1015, cs12b1018, msingh\}@iith.ac.in}}

\maketitle

\begin{abstract} 
A large amount of content is generated everyday in social media. 
One of the main goals of content creators is to spread their 
information to a large audience. There are many factors that affect
information spread, such as posting time, location, type of information, 
number of social connections, etc. In this paper, we look at the problem
of finding the best posting time(s) to get high content visibility. The posting
time is derived taking other factors into account, such as location, 
type of information, etc. In this paper, we do our analysis over Facebook
pages. We propose six posting schedules that can be used
for individual pages or group of pages with similar audience reaction profile. 
We perform our experiment on a Facebook pages dataset containing 0.3 
million posts, 10 million audience reactions.  Our best posting schedule can lead to
seven times more number of audience reactions compared to the average
number of audience reactions that users would get without following
any optimized posting schedule. We also present some interesting 
audience reaction patterns that we obtained through daily, weekly and monthly 
audience reaction analysis. 
\end{abstract}


\section{Introduction}
\label{sec:rtp_int}
Social media includes various web-based services that allow users 
to create and share the content with other users within their social network. 
A large amount of data is generated daily in social media. 
One of the main goals of a content creator is to spread the information to a 
large audience, and thereby receive a large number of audience reactions in the 
form of likes, comments, shares, etc.

The main obstacle in getting high information spread is that a 
post has a very short lifetime and within this short lifetime it has to 
compete with many other posts~\cite{asur2011trends,guille2013information}. In this paper, we use audience reaction as a measure to evaluate the information spread. Many factors affect 
audience reactions, such as posting time, content type, location,  
connection in social network, and so on. In this paper, we primarily look
at what is the effect of posting time on audience reactions? We propose techniques
to compute posting schedules that will lead to increase audience reactions.

In this paper, we use publicly accessible Facebook pages to create our dataset. 
Facebook pages are maintained by brands, businesses, 
organizations, etc., to inform customers about their products and services. There are
two types of social network relationships: {\it friend} relationship and 
{\it follower-following} relationship. Facebook pages use follower-following kind of relationship.  
Each page has admin(s) who create contents in the form of posts. Users can
follow the page and create reactions in the form of likes, comments, and shares. We 
call these users as audience. Posted content is broadcasted to the news feed of 
followers and it has to compete with many other contents to be at the top of the
followers' news feed. 

In social media, most of the audience reactions are received within first few 
hours of posting~\cite{wu2011says}. 
If a content is posted at a time when audience are not online or 
not interested in interacting with the content,
the content will not receive a large number of audience reactions. 
Facebook's News Feed algorithm~\cite{facebook:NewsFeed} 
rewards a post if it is getting a large number of audience reactions  
by increasing rank of the post. 
If the post appears at the top of the news feed of many users,
it would get more audience reactions and thereby becomes more
popular. 

Apart from looking at the ideal posting time for individual pages,
it would be interesting to characterize the pages into groups with
similar audience reaction profile. This will enable us to understand
what are the factors that determine audience reactions. 
Given there are millions of Facebook pages, 
creating page category and then computing the posting schedule
for the whole category will give higher statistical confidence 
while comparing the similarity and differences between various pages.
With this
characterization, we can also determine what would be the ideal posting
schedule for a new page which does not have enough audience 
interaction history.
Let us consider the following example task:
\begin{example} 
Consider a set of traffic related Facebook pages, where each page contains 
information about traffic updates for a particular city. Following are
some of the questions that we address in this paper:
\begin{enumerate}
\item What is the best time in a day that one should post about traffic updates 
to get maximum audience reactions?
\item Is there any difference in the audience reaction pattern over the week?
\item Are there typical periods during the year in which people tend 
to look more at traffic updates? 
\item How audience reaction pattern of traffic pages compare with other types of Facebook pages?
\end{enumerate}
\end{example}
\noindent Our key contributions are as follows:
\begin{itemize}
\item We analyze post-to-reaction behavior of Facebook pages. We show that
84\% of the audience reactions are received within 24 hours after posting. 
\item We identify top features that affect audience reactions and use these features
to categorize pages into groups with similar audience reaction profile. 
\item We propose six posting schedules for individual pages and groups of similar pages. 
\item We evaluate our algorithms on a dataset  
with 0.3 million posts and 10 million audience reactions.
Our best posting schedule can lead to
seven times more number of audience reactions compared to the average
number of audience reactions that one would get without following
any optimized posting schedule. 
\end{itemize}

The rest of the paper is organized as follows. We formally define the problem of finding the right time to post to maximize the visibility of content in Section~\ref{sec:rtp_plm}. Section~\ref{sec:rtp_arb} presents the audience reaction behaviour on Facebook pages. Section~\ref{sec:rtp_cfp} discuss the categorization methods.  Section~\ref{sec:rtp_sd} introduces the algorithm for schedule derivation. We proceed by describing schedule evaluations in Section~\ref{sec:rtp_se}. We briefly go through the related work in Section~\ref{sec:rtp_rw} and conclude our work in Section~\ref{sec:rtp_cnf}.

\section{Problem Formulation}
\label{sec:rtp_plm}
In this section, we present the problem definition 
and details about the used dataset.

\subsection{Problem Definition}
The problem of finding the right time to post can be defined in 
terms of the following sequence of sub-problems:

\noindent {\bf Problem 1 (Schedule for a Facebook page)}: {\it Given 
a Facebook page ${P}$, find a set of time-interval(s) ${T_P}$ 
such that if a post $p \in {P}$ is posted during any time-interval
$t_k \in {T_P}$, the post $p$ is likely to get high visibility, 
which is measured using the number of audience reactions received on $p$.
}

Problem 1 is the right time to create a post for a single Facebook page. 
If a post is created according to the proposed schedule ${T_P}$, 
it would get more audience reactions. 
According to Facebook's {\em News Feed} algorithm~\cite{facebook:NewsFeed}, 
if a post is getting a large number of audience reactions, the post will 
be given a chance to appear on top of the news feed of more number of users, 
thereby further increasing its likelihood to get high audience reactions. 
The schedule can be derived by using the posting behaviour 
of page admins (pages) or the reaction behaviour of audience. We state
these two problems below. 

\noindent {\bf Problem 1.1 (Frequent Posting Schedule)}:{ \it 
Given a Facebook page ${P}$ or a page category ${C}$, 
and the post creation profile $M$, find the frequent
posting schedule $S^{fp}$ for the page ${P}$ 
or the category ${C}$.}

Admins of Facebook pages post a content at the time they receive the content
(or just follow a certain personal schedule to post their contents).
Although many admins may not be aware of when they should post
to get maximum audience reactions, some expert admins with
knowledge of social media post ranking might have an intuition of when
they should post to get maximum audience reactions. They might realize
this by trying out various posting schedules. 
Thus, our first problem is based on the most
frequent posting schedule~(\textit{category} is defined later in Problem 2). 

Frequent posting schedule can be of three types: {\em aggregated}, 
{\em category specific} and {\em weighted category specific} denoted as 
$S^{afp}$, $S^{cfp}$, and $S^{wcfp}$ respectively. 
The aggregated schedule is the common schedule that can be
used by all the pages. Categorized schedule is the customized schedule for the categories, and 
it is the best schedule for all the pages in a given category. 
Within a given category, all the pages may not have the same importance.
Weighted category specific schedule is derived by giving higher
weight to more important pages within the category.

\noindent {\bf Problem 1.2 (Frequent Reaction Schedule)}:{ \it 
Given a Facebook page ${P}$ or a page category ${C}$, 
and the audience reaction profile $R$, find the frequent
reaction schedule $S^{fr}$ for the page ${P}$ 
or the category ${C}$. }

Since our goal is to maximize the number of audience reactions,
the frequent reaction based schedule is derived by analyzing the posting
timings that lead to high audience reaction. Frequent reaction
schedules are also of three types: {\em aggregated}, 
{\em category specific} and {\em weighted category specific} denoted as 
$S^{afr}$, $S^{cfr}$, and $S^{wcfr}$ respectively.

\noindent{\bf Problem 2 (Facebook page Categories)}: {\it 
Given a set of Facebook pages $\mathcal{P}$, a set of reaction determining features ${F_R}$, 
categorize the pages in $\mathcal{P}$ into $r$ categories $\{C_1,C_2,..,C_r\}$
such that similarity between reaction profile is high for pages within a category
and low across categories.}

Each Facebook page has a unique pattern of audience reaction. 
The pattern is not same for all the pages. Analyzing these
reactions will help the page admins to get a deeper insight into
their pages. For example, two e-commerce websites may have the different
type of audience reaction patterns, even though they may be from the same location or the similar type of organization. By categorizing
pages into categories with similar audience reaction profile, we 
can understand what are the different types of audience reaction 
profile? What are the factors that cause one page to get a certain type
of audience reaction profile? If an organization wants its page to attain popularity
similar to some other organization, what are the factors the organization
should focus on to achieve that level of popularity? All these questions
can be answered by looking at category-wise reaction behavior.
 
\subsection{Dataset}
\label{sec:rtp_dataset}
We do our analysis on publicly accessible Facebook pages having a large number of audience. 
We obtain the dataset using the Facebook Graph API\footnote{https://developers.facebook.com/docs/graph-api} in a similar way as described by Weaver et al.~\cite{weaver2013facebook}. 
Each page has a profile page that contains posts created by page (posts created by the admin of page) and the reactions received on posts from the audience.
Each page has a label (organization name) and 
a set of attributes (features). These attributes can vary across pages. 
A page can have attributes such as the number of fans (users who liked the page), the number of people talking about the page, type of the page, organization name, post creation time, reaction time, etc.

Audience can react on the posts created by Facebook pages 
in the form of like, comment and share. 
Reactions consist of a textual comment and 
a unary rating score in form likes and shares. 
As an audience member reads a post, she can 
optionally create a reaction to the post created by Facebook Page.
Each audience member can 
contribute one or multiple reactions to a post. 
Audience are allowed to update previous reactions and 
add new reactions on the reacted posts. 
Since we could only access timestamp for comments, we use comments 
as the reaction and the time of comments creation as the reaction timestamp. 
Comments can be used to implicitly measure the interest generated by a post~\cite{mazloom2016multimodal,tatar2014popularity}. 
We extract the data of 100 Facebook pages from the same location that 
includes 5 different categories namely, e-commerce, traffic, telecommunication, hospital, and politician.
Each of these categories contains the same number of pages to maintain homogeneity in audience reactions across the categories.

\begin{table}[!h]
\begin{center}
\begin{tabular}{|c|l|}
\hline
\hline
\thead{\bf Notation} & \thead{\bf Number}\\
\hline
\hline
 ${R}$: Reactions & 10 million \\
 \hline
 ${M}$: Posts/Messages & 0.3 million\\
 \hline
${Y}$: Years  & 5 years \\
 \hline
 ${N}$: Number of pages  & 100 \\
 \hline 
\end{tabular} 
\end{center}
\caption{Dataset Statistics} 
\vspace{-0.1in}
\label{table:data}
\end{table}
\vspace{-0.1in}
 
As can be seen in Table~\ref{table:data}, our collected dataset contains 0.3 million posts and 10 million
reactions that were created in 5 years (2011-2015). 
As the dataset contains many unimportant and noisy words, we pre-process the data using text-processing techniques~\cite{perkins2014python} such as stop-word removal, stemming, lemmatization, etc.
We remove stop words from posts and 
comments as these words do not contain important significance to be used in the analysis. 
We also perform stemming and lemmatization to reduce inflected or derived words to their root 
forms.

\section{Audience Reaction Analysis}
\label{sec:rtp_arb}
In this section, we look at the user dynamics in Facebook pages. 
We analyze the time delay between when a post is created and when the audience
react to it. We also show different types of audience reaction that pages receive. 

\subsection{Post to Reaction Time Analysis}
\begin{figure}[h]
    \centering
   \fbox{\includegraphics[width=6.5cm, height=4.5cm]{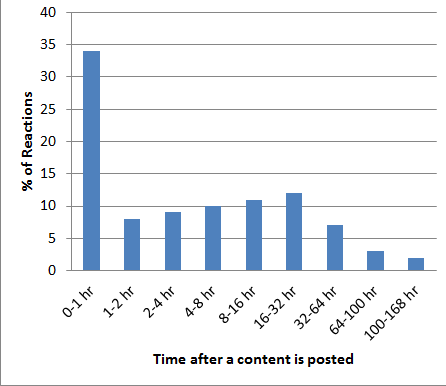}}
    \caption{Distribution of Reactions on a Post}
    \label{fig:rt1}
\end{figure}

There is some time lag in post creation and audience reaction 
time~\cite{asur2011trends,wu2011says}. It is important to study this time delay as
some of the important features used to find the right time to post are derived 
from this time delay.
Typically, a post receives 97\% of its total audience reactions within the first week of its posting. So, we consider timespan of one week 
to analyze post-to-reaction delay. Figure~\ref{fig:rt1} shows the 
distribution of audience reactions over a period of a week.

We observe in Figure~\ref{fig:rt1} that a post receives around 34\% of its
total reactions within the $1^{st}$ hour of its posting, and 84\% of reactions within a day.
The lifespan of a post is very short,
typically few hours and 
if it is not posted at the right time, it may not get high audience reactions. 
So it becomes important for Facebook pages to 
choose a right time of the day to post a content.
A Facebook page can post a limited number of posts per day/week. If a page creates fewer posts, 
it will not engage audience enough for them to maintain a social connection with the 
page and the page will lose engagement. On the other hand, if a page 
creates a lot of posts, it will typically lose engagement. 
So, it is important to know the right time~(daily, weekly, monthly) to 
create a post in Facebook page. This is the motivation for our 
proposed problem to find the right time to post to get maximum
content visibility.

\subsection{Audience Reaction Behavior Analysis}
We present audience reaction behavior profile of some real world Facebook pages to 
understand the diversity of audience reaction pattern. We look at individual pages from
politics, e-commerce, telecommunication, traffic, and hospital. 

\begin{figure}[!h]
    \centering
  \fbox{\includegraphics[width=10cm, height=5cm] {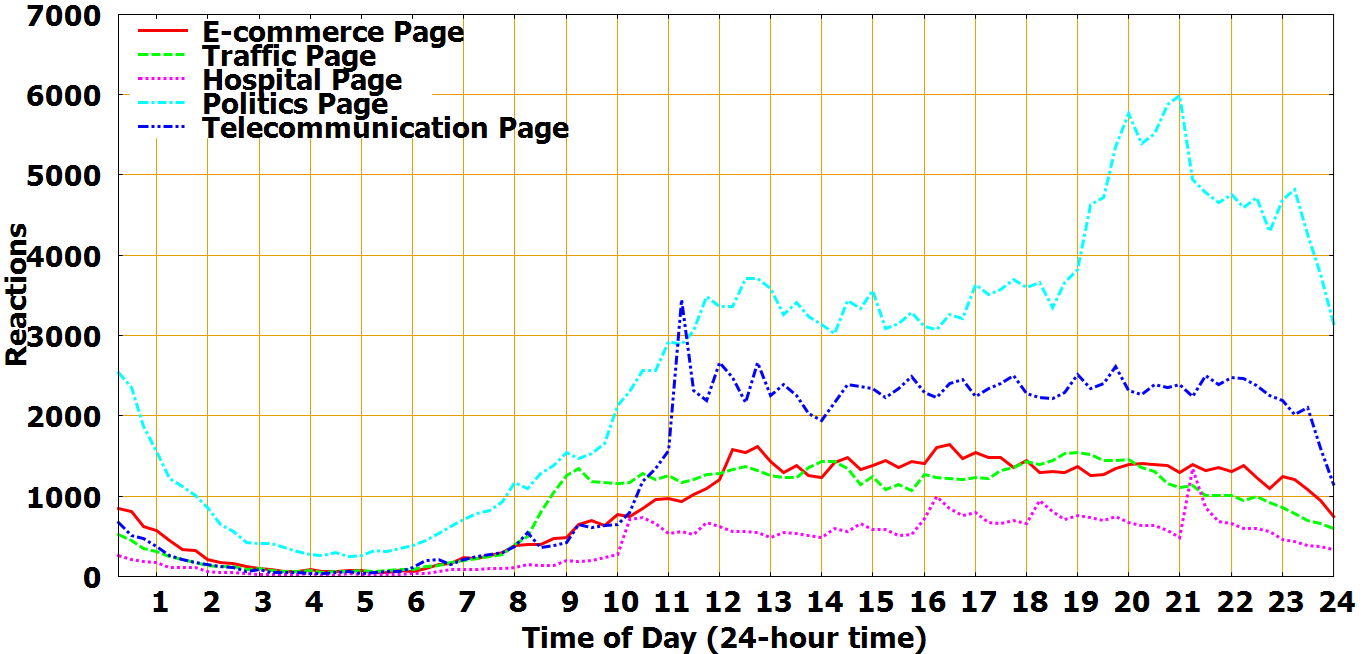}}
    \caption{Audience Reaction Behavior}
    \label{fig:arr}
\end{figure}

As can be seen in 
Figure~\ref{fig:arr}, the audience reaction behavior vary across time and pages. 
Some pages have one or more peaks per day. Some pages have a uniform peak 
throughout the day. 
The page maintained by a politician, receive peak audience reaction between 8:00 pm - 10:00 pm.
Audience reaction is much less during rest of the day. For e-commerce and telecommunication related 
pages the peak is around 11 am, and then it decreases a bit for the rest of the day.
It indicates that the audience reactions also depend on the content and 
characteristics of the page~\cite{biswas1994mood,reis2015breaking,esiyok2014users,yan2012better}. 
We give more detailed results on audience reaction analysis in Evaluation Section~\ref{sec:rtp_se}.

\section{Categorization of Pages}
\label{sec:rtp_cfp}
In this section, we give a solution to Problem 2. We present the reaction determining features and describe the method of feature processing, page categorization. 
 
\subsection{Reaction Determining Features}
\label{sec:df}
To find features that affect audience reactions, we create 35 features. 
We use wrapper based feature selection to select the
top reaction determining features. The features can be divided into 
following three types:
  \vspace{-0.11in}
\subsubsection{Page centric features} These are the features about the pages and signify popularity of the pages. 
Example features include the number of fans (those who have liked the page), the fan growth rate, the number of people who have created a story about the page on Facebook, and the number of posts per day.  

  \vspace{-0.11in}
\subsubsection{Content centric features} These are the features about the page content.
Example features include type of the page (described in Section~\ref{sec:rtp_fp}); average number of likes,
comments and shares for the whole page; average likes, comments and shares for different
types of contents, such as Photos, Links, Videos, and average post length.  

  \vspace{-0.11in}
\subsubsection{Reaction centric features} These are the features about audience reaction.
Example features include the average number of audience reactions received within various
time intervals after the post is created, such as 0-1 hrs, 1-2 hrs, 2-4 hrs, 4-8 hrs, 
8-16 hrs, and 16-32 hrs; the average number of audience reactions received during various 
day intervals, such as 12:00 am - 4:00 am, 4:00 am - 8:00 am, 8:00 am - 12:00 pm, and so on. 
These features also include the average number of reactions received on days of a week and months of a year.

\subsection{Feature Pre-processing}
\label{sec:rtp_fp}
\vspace{-0.05in}
We perform various pre-processing for the above
features, such as correct the time zone, correct the type 
of page, convert continuous valued attributes to discrete valued attributes.   
We extract the timestamp associated with each post and reaction. Graph API provides the time in Greenwich Mean Time (GMT) format; we convert it into regional time-zone. 

Admins of Facebook pages create the label (or type) for their 
pages, and they name it based on the domain of the page/organization. 
There are six primary labels provided by Facebook for pages namely, “Local Business or Place”, “Company Organization or Institution”, “Brand or Product”, “Artist, Band or Public Figure”, “Entertainment”, “Cause or Community”. Each of these labels includes multiple sub-labels such as “Brand or Product” includes “website”, “electronics”, “product/service”, etc. Each page admin has to select one of these labels for their page. There are inconsistency between admins on how they select labels. For example, one e-commerce page is labelled as “Retail Company” and the other is labelled as “Website”. 
We use Nearest Neighbor algorithm~\cite{garcia2009k} to label pages in a consistent manner, as page label is one of the most important factors in our posting schedule analysis. We use topic modeling to represent the pages in terms of topics, and then use cosine similarity of their topic probability to compute similarity between pages. For each page, we find its k-nearest neighbor pages. We then use majority label from these k neighbors to correct the page label. If the page label is labelled correctly, then majority will also have the same label. If it is not the most appropriate label, then it will differ from the majority and we correct it by assigning the majority label.
Since organizations from
similar domain post similar type of information, this technique
can give all the pages of same domain the most common
label used in that domain. 

In order to characterize the pages based on these feature attributes, we convert these continuous attributes to discrete attributes. 
We apply entropy based data discretization~\cite{fayyad1992handling} 
method to convert features in discrete attributes because 
most of the unsupervised data discretization methods require some parameter $k$ such as number of bins. 
Entropy based method search through all possible values of $k$ and 
capture inter-dependencies in features. 

 \vspace{-0.17in}
\subsection{Categorization}
 \vspace{-0.05in}
We use clustering to group the pages with similar audience reaction.
We use wrapper method~\cite{kohavi1997wrappers} to select the features that are relevant
for audience reaction.  It considers selection of a subset of features as a search problem, 
where different combinations of features are used, evaluated and compared to other combinations.
In the wrapper method, we use Multinomial Naive Bayes classifier~\cite{mccallum1998comparison} for classification. 
To create the base classes of Multinomial Naive Bayes, we use
$k$-medoid clustering algorithm over the pages, where $k$ is chosen using elbow method~\cite{kodinariya2013review}. 
We define the similarity (refer to Equation~\ref{eq:rtp_co})  
between two pages $P_i$ and $P_j$ in $k$-medoid as the similarity between their reaction profile $R_k(P_i)$ and $R_k(P_j)$ (reaction profile is defined in Section~\ref{sec:as}). 
We use $k$-medoid algorithm instead of $k$-means algorithm because of 
its robustness to outliers as compared to $k$-means. Moreover, it uses representative 
objects as cluster centers instead of taking the mean value of the objects as a
cluster center. 
The top three obtained features are the \textit{reaction within first one hour}, 
\textit{number of posts posted by the page per day} and \textit{type of the page} in increasing order of usefulness for the categorization.
We cluster the pages using the top three reaction determining features as 
these three features are able to classify the pages into right category with the highest accuracy (90.3\%)
and increasing the number of features does not make a significant change in accuracy. 
The similarity in audience reaction within a category is high and across the categories is low when we use theses three features for categorization (as shown in Section~\ref{sec:eoc}). 
 \vspace{-0.1in}

\section{Schedule Derivation}
\label{sec:rtp_sd}
In this section, we give a solution to Problem 1. First, we describe notations used 
in schedule derivation and later present six ways to compute 
posting schedule. The first two schedules are generic
schedules that are applicable for all pages, whereas the last four 
schedules are category specific.

\begin{table*}[!h]
\begin{center}
\begin{tabular}{|l|l|}
 \hline\hline
 \bf Symbol & \bf Description \\ [0.25ex] 
  \hline\hline
  $P$ & a given set of Facebook pages \\
 \hline
 $C_i$ & a set of similar Facebook pages, $C_i \subseteq P$\\
 \hline
 $Y_s$, $d$	& $Y_s$ is the base year in the dataset, $d$ is the total number of years\\
 \hline
  $t_k$	& a time bucket of size 15 minute \\ 
 \hline
  $r_k(P_x, Y_j)$ & reaction profile vector of page $P_x$ in $Y_j^{\text th}$ year\\
 \hline
  $R_k(P_x)$ & cumulative reaction profile vector of page $P_x$ across $d$ years \\
 \hline
  $m_k(P_x, Y_j)$ & posting profile vector of page $P_x$ in $Y_j^{\text th}$ year\\
 \hline
  $M_k(P_x)$ & cumulative posting profile vector of a page $P_x$ across $d$ years \\
 \hline
  $\gamma^r{(P_x)}$ & total number of reactions received in page $P_x$ in $d$ years\\
 \hline
  $\gamma^m{(P_x)}$ & total number of posts created by page $P_x$ in $d$ years\\
 \hline
  $\rho^r{(C_i)}$ & total number of reactions received in category $C_i$ in $d$ years\\
 \hline
  $\rho^m{(C_i)}$ & total number of posts created by category $C_i$ in $d$ years\\
 \hline
 $W^m(P_x)$ & fraction of posts created by page $P_x$ within its own category\\
 \hline
  $W^r(P_x)$ & fraction of reactions received in page $P_x$ within its own category\\
 \hline
   $\delta(C_i, k)$ & reaction per post for category $C_i$ in $k^{\text{th}}$ bucket across $d$ years\\
 \hline
   $\omega(C_i)$ & aggregated reaction per post for category $C_i$ in all buckets across $d$ years\\
 \hline
\end{tabular}%
\caption{Notations}
\label{table:crt}
\end{center}
\end{table*}

Let's assume we have data from $d$ years \{$Y_s,Y_s+1,.....,Y_s+d$\}. We divide a day into 96 discrete buckets \{$t_1,t_2,...,t_{96}$\}, with each bucket of size 15 minutes as the bucket can capture essential reactions (as shown in Figure~\ref{fig:rt1}). By dividing a day time into small size of 96 buckets, we are able to determine right time (or bucket) more precisely.
The first bucket $t_1$ is from 24:00 hrs to 00:15 hrs.
We aggregate actions in the same time bucket from multiple years
to ensure that our derived results are reliable. 
We consider two types of actions: creation (posting) and reaction. We
denote posting and reaction profile for a given time bucket $t_k$, 
page $P_z$, and year $Y_j$ as $m_k(P_z, Y_j)$ and
$r_k(P_z, Y_j)$ respectively. 
$m_k(P_z, Y_j)$ is the aggregated number of posts created by page $P_z$ at $Y_j^{\text{th}}$ year (all days of year $Y_j$) in the time bucket $t_k$. 
For each bucket $t_k$, $m_k(P_z, Y_j)$ is computed by counting the number of posts created by page $P_z$ in the time bucket $t_k$ over the year $Y_j$. 
$r_k(P_z, Y_j)$ is the aggregated number of reactions received in page $P_z$ at $Y_j^{\text{th}}$ year in the time bucket $t_k$. 
For each bucket $t_k$, $r_k(P_z, Y_j)$ is computed by adding all the reactions received by page $P_z$ in the time bucket $t_k$ over the year $Y_j$. 
We use these two profiles to compute the schedules. 

\subsection{Aggregated Schedules}
\label{sec:as}
We present two generic schedules,
which are common for all the pages. The first schedule ($S_k^{afp}(P)$) is based
on the \underline{a}ggregated \underline{f}requent \underline{p}osting behavior and the second
schedule ($S_k^{afr}(P)$) is based on \underline{a}ggregated \underline{f}requent
\underline{r}eaction behavior of all the pages.

Aggregated frequent posting schedule ($S_k^{afp}(P)$) is generated by using cumulative posting profile vector $M_k(P_z)$. 
For each time bucket $t_k$, $M_k(P_z)$ is the total number of posts created by page $P_z$ in time bucket $t_k$ across $d$ years. $M_k(P_z)$ is computed by aggregating the posting profile vector $m_k (P_z, Y_j)$ of page $P_z$ across $d$ years as follows:
\vspace{-0.01in}
\begin{equation}
M_k(P_z) =  \sum_{j=y_s}^{y_s+d} m_k (P_z, Y_j)    
\end{equation}
\vspace{-0.1in}

$S_k^{afp}(P)$ is a fraction of total number of posts created by all the pages in the $t_k^{th}$ bucket. It is computed as follows: 
\begin{equation}
\label{eq:safp}
S_k^{afp}(P)=  \dfrac{\sum_{z=1}^N M_k(P_z) }{ \sum_{z=1}^N \sum_{k=1}^{96}M_k(P_z)  }
\end{equation}
where  $P_z \in P$ and $S_k^{afp}(P)$ is the fraction of total posts created by pages in $k^{\text{th}}$ bucket, which is also defined as the probability of creating a post by pages in $k^{\text{th}}$ bucket.

Similarly, aggregated frequent reaction schedule ($S_k^{afr}(P)$) is generated by using cumulative reaction profile vector $R_k(P_z)$. 
For each time bucket $t_k$, $R_k(P_z)$ is the total number of reactions received by the page $P_z$ in the time bucket $t_k$ across $d$ years.
$R_k(P_z)$ is computed by aggregating the reaction profile vector $r_k (P_z, Y_j)$ of page $P_z$ across $d$ years as follows:

\vspace{-0.15in}
\begin{equation}
R_k(P_z) = \sum_{j=y_s}^{y_s+d} r_k(P_z, Y_j)
\end{equation}
\vspace{-0.05in}

$S_k^{afr}(P)$ is a fraction of total number of reactions received by all the pages in the $t_k^{th}$ bucket. It is computed as follows: 
\begin{equation}
\label{eq:safr}
S_k^{afr}(P)=  \dfrac{\sum_{z=1}^N R_k(P_z) }{ \sum_{z=1}^N \sum_{k=1}^{96}R_k(P_z)  }
\end{equation}
where $S_k^{afr}(P)$ is also defined as the probability of receiving audience reaction on pages in the $k^{\text{th}}$ bucket. 
Now, we rank the buckets in decreasing order of $S_k^{afr}(P)$, $S_k^{afp}(P)$ with 
the first bucket being the best and the last one being the worst time to post according to these schedules respectively.

\subsection{Categorized Schedules}
\label{sec:cs}
As each category has different reaction behavior compared to other categories, 
we generate customized schedule for each category
of Facebook pages. 
We derive two customized schedules for categories of Facebook pages, 
namely \textit{categorized frequent posting schedule} and \textit{categorized frequent reaction schedule}.

Categorized frequent posting schedule $S_k^{cfp} (C_i)$ is computed based on  
number of posts created by category $C_i$ in time bucket $t_k$, and total number of posts created by category $C_i$ in all the buckets as follows: 
\begin{equation}
S_k^{cfp} (C_i) = \dfrac{\sum_{x=1}^{|C_i|} M_k(P_x)}  {\sum_{k=1}^{96} \sum_{x=1}^{|C_i|} M_k(P_x)}
\end{equation}
where $P_x \in {C_i} $, $M_k(P_x)$ is the cumulative posting profile vector of page $P_x$ and $|C_i|$ is the total number of 
pages in category $C_i$. $S_k^{cfp} (C_i)$ is the fraction of total posts posted by category $C_i$ in $k^{\text{th}}$ bucket, which is also defined as the probability of creating a post by category $C_i$ in $k^{\text{th}}$ bucket. 
Similarly, categorized frequent reaction schedule ($S_k^{cfr}(C_i)$) is computed as follows:
\begin{equation}
S_k^{cfr}(C_i) = \dfrac{\sum_{x=1}^{|C_i|}  R_k(P_x)}  {\sum_{k=1}^{96} \sum_{x=1}^{|C_i|}  R_k(P_x)}
\end{equation}
where $R_k(P_x)$ is the cumulative reaction profile vector of page $P_x$. $S_k^{cfr} (C_i)$ is the fraction of total reactions received on category $C_i$ at $k^{th}$ bucket, which is also defined as the probability of receiving audience reaction on category $C_i$ in $k^{\text{th}}$ bucket.

We rank the buckets in decreasing order of $S_k^{cfp} (C_i)$ and $S_k^{cfr}(C_i)$. We pick first few buckets from both the schedules which are the 
right time to post for a category $C_i$ according to these schedules. We compute categorized schedules for all the categories by following the same procedure. 
First time bucket of ranked schedules is the 
best time to post for category $C_i$ in order to maximize content visibility.

\subsection{Weighted Categorized Schedules}
\label{sec:wcs}
We derive the weighted categorized schedules by 
assigning weight to the pages of categories based on their importance. 
Some of the pages receive a large number of audience reactions and some of the pages post a large number of posts compared to other pages.
To maintain homogeneity of actions and audience reactions across all pages in a category, 
we use weight factor ($W^r(P_x)$, $W^m(P_x)$) in computation of the schedules. 
Weight signifies the importance of each page in its category. 
It is computed by using two parameters $\gamma$ and $\rho$ as follows:
\begin{equation}
\gamma^r(P_x) = \sum_{k=1}^{96} R_k(P_x)  
\end{equation}

\begin{equation}
\rho^r(C_i)= \sum_{x=1}^{|C_i|} \gamma^r(P_x)
\end{equation}

\begin{equation}
W^r(P_x) = \dfrac{\gamma^r(P_x)}{\rho^r(C_i) }
\end{equation}
where $\gamma^r(P_x)$ is the total number of reactions received by a page $P_x$ and $\rho^r(C_i)$ is the total number of reactions received by a category $C_i$ (all the pages of the category). 
Similarly, $\gamma^m(P_x)$, $\rho^m(C_i)$, and $W^m(P_x)$ are computed using cumulative posting profile vector ($M_k(P_x)$). 
Weighted categorized frequent posting schedule $S_k^{wcfp}(C_i)$ for category ($C_i$) is computed as follows: 
\begin{equation}
S_k^{wcfp}(C_i) = \dfrac{\sum_{x=1}^{|C_i|} W^m(P_x) \times M_k(P_x)} { \rho^m(C_i) }
\end{equation}
where $S_k^{wcfp}(C_i)$ computes the probability of creating a post by a category $C_i$ at the $k^{\text{th}}$ bucket. 
Now, we compute weighted categorized frequent reaction schedule $S_k^{wcfr}(C_i)$ 
for a category ($C_i$) as follows:
\begin{equation}
S_k^{wcfr}(C_i) = \dfrac{\sum_{x=1}^{|C_i|} W^r(P_x) \times R_k(P_x)} { \rho^r(C_i) }
\end{equation}
where $S_k^{wcfr}(C_i)$ computes the probability of receiving audience reaction on category $C_i$ in $k^{\text{th}}$ bucket.

Weighted categorized schedule is similar to categorized schedule, the only difference is that weighted categorized schedule is computed by assigning a weight to each page of a category based on its importance in that category.
We rank the buckets in decreasing order of $S_k^{wcfp} (C_i)$ and $S_k^{wcfr}(C_i)$ for 
all the categories. We pick first few buckets from both the schedules which are the 
right time to post for a category $C_i$ according to these schedules. 
We compute weighted categorized schedules for all the categories.

\section{Evaluations}
\label{sec:rtp_se}

In this section, we evaluate our proposed schedules, page categorization technique and present the audience reaction behaviour over time. We also discuss how the audience engagement varies with the type of post content. 

\subsection{Evaluation Metrics}
We use \textit{reaction gain} to evaluate the schedules
and \textit{correlation} to evaluate the quality of
our categorization function.
 \vspace{-0.11in}
\subsubsection{Reaction Gain}
\label{rtp:subsec_rg}
Reaction gain metric is used to compute the performance of proposed schedules. It measures the change in reactions received in a particular time bucket, compared
to the average reactions per post. Before computing the reaction gain for a schedule ($S$), we first rank the time buckets of schedule ($S$) over a period
of 24 hours and compute two parameters: 
{\em reaction per post} $(\delta)$ and {\em aggregated reaction per post} $(\omega)$. 
Reaction per post $(\delta)$ is the total number of reactions received on  
pages within category $C_i$ at time bucket $t_k$ in $d$ years divided by the total number of posts 
created at time bucket $t_k$ by category $C_i$ in $d$ years. For the $k^{\text{th}}$ rank bucket as per schedule ($S$) of category $C_i$, {\em reaction per post} $(\delta)$ is computed as follow: 
\begin{equation}
\delta(C_i,k) =\dfrac{R_k(C_i)} {M_k(C_i)}
\end{equation}
where $R_k(C_i)$ and $M_k(C_i)$ are the cumulative reaction profile vector and cumulative posting profile vector for the category $C_i$ respectively. 
$R_k(C_i)$ and $M_k(C_i)$ are computed by aggregating the cumulative reaction profile vectors, cumulative posting profile vectors of all the pages in its own category respectively. 

Aggregated reaction per post $(\omega)$ is the total number of reactions 
received on pages of category $C_i$ divided by 
the total number of posts created by pages of category $C_i$.
\begin{equation}
\omega(C_i) = \dfrac{\sum_{k=1}^{96}R_k(C_i)} {\sum_{k=1}^{96} M_k(C_i)}
\end{equation}

Now, reaction gain (RG) for time bucket $t_k$ and category $C_i$ is defined as: 
\begin{equation}
RG(C_i,k) =\dfrac{\delta(C_i, k)} {\omega(C_i)}
\end{equation}
where $RG(C_i, k)$ signifies the increase or decrease in reactions received by the category $C_i$ when it posts in time bucket $t_k$, compared to the average reactions per post it receives. 

Similarly, we compute the reaction gain ($RG(P,k)$) for the aggregated schedules by using $\delta(P,k)$, $\omega(P)$,
$R_k(P)$, and $M_k(P)$. $R_k(P)$ and $M_k(P)$ are determined by aggregating the cumulative reaction profile vector and cumulative posting profile vector of all the pages respectively. 
Next, we compute average reaction gain for the categorized and weighted categorized schedules as these schedules contain multiple categories. 
\begin{equation}
RG_{avg}(k) =  \dfrac{\sum_{i=1}^r RG(C_i,k)} {r}
\end{equation}
where average reaction gain ($RG_{avg}(k)$) 
for $k^{\text{th}}$ time bucket is the average of $RG(C_i,k)$ across all the $r$ categories. 
We use $RG_{avg}(k)$, $RG(P,k)$ to evaluate the performance of categorized schedules and aggregated schedules respectively.

\subsubsection{Correlation}
We use correlation metric to evaluate the effectiveness of the categorization method. 
We compute correlation across the categories by using the cumulative reaction profile vector of categories as follows:

\begin{equation}
Co(C_i, C_s)=\dfrac{\sum_{k=1}^{96} (   R_k(C_i) -\bar{R}(C_i) ) *  (   R_k(C_s) -\bar{R}(C_s) ) }{\sqrt{\sum_{k=1}^{96} (R_k(C_i) -\bar{R}(C_i))^2 }* \sqrt{\sum_{k=1}^{96} (R_k(C_s) -\bar{R}(C_s))^2 }} 
\label{eq:rtp_co}
\end{equation}

where $C_i$ and $C_s$ are two different categories. $R_k(C_i)$ is the cumulative reaction profile vector (audience reaction) of category $C_i$ in $k^{\text{th}}$ bucket and $\overline{R}(C_i)$ is the 
average audience reaction of category $C_i$. 

Similarly, we use the cumulative reaction profile vectors of categories of pages ($R_k(P_x)$) to compute the correlation within the category.
We determine the correlation within the category by taking the average of correlation computed between each pair of the pages which belong to the same category.

\subsection{Effect of Schedule}
\label{sec:eos}
We evaluate our proposed six schedules using reaction gain metric defined in Section~\ref{rtp:subsec_rg}. 
As there are no previous baselines on best time to post for Facebook pages, we consider the first two generic schedules, namely aggregated frequent posting schedule and aggregated frequent reaction schedule as baseline schedules. 
We compute the average reaction gain for all the categorized schedules, aggregated schedules and 
pick the top-$30$ time buckets. 

\begin{figure}[h]
    \centering
   \includegraphics[width=10cm, height=5cm]{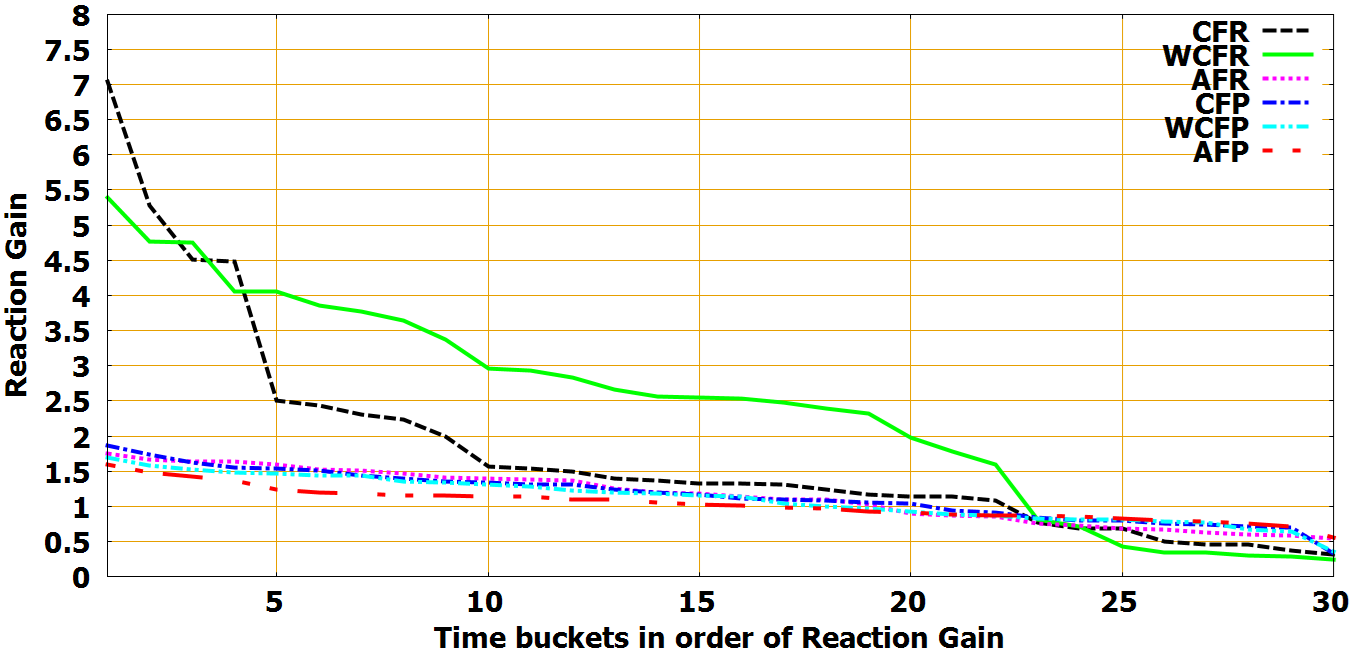}
    \caption{ Reaction Gain  }
    \label{fig4}
\end{figure}

We observe in Figure~\ref{fig4} that all the posting based 
schedules, such as $S^{AFP}$, $S^{CFP}$, and $S^{WCFP}$ have
reaction gain less than 2.0 even in their top bucket and
their overall performance is also not as good as reaction based schedules. 
The reason is that most page admins do not know what 
is the right time to post a content. They may not 
be even aware of the fact that they can get better audience reaction
by just choosing a better time for posting.

On the other hand, reaction based schedules perform 
far better compared to posting based schedules. 
It is also observed that category-wise schedules perform better than 
aggregated schedules (baseline schedules). 
Reaction gain of categorized frequent reaction schedule 
($S^{CFR}$) is highest (i.e., seven times better) in its top bucket. 
Weighted categorized frequent reaction ($S^{WCFR}$) 
schedule shows a reaction gain of 5.4 in the top bucket. 
$S^{WCFR}$ performs better than the $S^{CFR}$
for all the buckets except the first two buckets. 
The reason could be that $S^{CFR}$ is biased towards those buckets which 
receive a large number of audience reactions.
If a page or category receives a large number of audience reactions in few buckets, 
it reflects high reaction gain in these buckets.
However, $S^{WCFR}$ is a normalized schedule and it does not 
show high reaction gain if few buckets receive 
high audience reaction.

\subsection{Effectiveness of Categorization}
\label{sec:eoc}
We compute the correlation within and across categories to show the effectiveness of our categorization method. 
Let's consider five categories: C1, C2, C3, C4 and C5. We label these
categories using the type of most frequent pages in that category.
With this, the categories C1, C2, C3, C4, C5 represent e-commerce, telecommunication, hospital, politics, traffic respectively.
We consider two ways of doing categorization: using single feature and
using multiple features. From the top reaction determining features, we select
the best feature for single feature case. In multiple feature case, we consider
all the top reaction determining features.

\begin{table}[H]
\parbox{.48\textwidth}{
\centering
\begin{tabular}{|c|l|l|}
\hline
\hline
\thead{\bf Categories} & \thead{\bf Single\\ \bf Feature} & \thead{\bf Multiple\\ \bf Feature}\\
 \hline\hline
 C1 \&  C2 & 0.547 & 0.503  \\ 
 \hline
  C1 \& C3 & 0.392 & 0.341 \\
 \hline
  C1 \& C4 &0.418 & 0.367 \\
 \hline
 C1 \& C5 & 0.519 & 0.470 \\
 \hline
 C2 \& C3 &0.403 &  0.378\\ 
 \hline
 C2 \& C4 & 0.353 & 0.302 \\
 \hline
 C2 \& C5 & 0.510 & 0.473 \\
 \hline
C3 \& C4 & 0.351 & 0.305 \\
 \hline
 C3 \& C5 & 0.448 & 0.416 \\
 \hline
 C4 \& C5 & 0.440 & 0.419 \\[1ex] 
 \hline
\end{tabular}
\caption{Correlation across the categories}
\label{table:2}
}
\hfill
\parbox{.48\textwidth}{
\centering
\begin{tabular}{|c|l|l|}
\hline
\hline
\thead{\bf Category} & \thead{\bf Single\\ \bf Feature} & \thead{\bf Multiple\\ \bf Features}\\
 \hline\hline
 C1 & 0.634 & 0.768\\ 
 \hline
 C2 & 0.703 & 0.848\\
 \hline
 C3 & 0.621 & 0.702\\
 \hline
 C4 & 0.650 & 0.771\\
 \hline
 C5 & 0.672 & 0.778\\ [1ex] 
 \hline
\end{tabular}
\caption{Correlation within the category}
\label{table:3}
}
\end{table}

We show across and within category
correlation in Table~\ref{table:2} and~\ref{table:3} respectively for both types 
of categorization. Ideally, we would want within category correlation high and 
across category correlation low. In case of single feature case, we find that 
within and across category correlation is almost same. However, in case 
of multi-feature categorization, there is a large difference between within 
and across category correlation. 
These results indicate that our categorization function is able to categorize the pages effectively using multiple features. A new page that doesn't have enough reactions, can use this analysis to determine its right category and can post the accordingly (as described in Section~\ref{sec:trp_ta}) to get a large number of audience reactions. 
For ease of presentation, in rest of the paper, 
we refer the categories as e-commerce, politicians, etc. Each of these categories contains the same number of pages to maintain homogeneity in audience reaction across the categories.

\subsection{Trend Analysis}
\label{sec:trp_ta}
We present some examples of audience
reaction patterns which is observed in daily, weekly and monthly analysis.

\subsubsection{Daily Analysis}
\begin{figure}[h]
\centering
  \includegraphics[width=9cm, height=5cm]{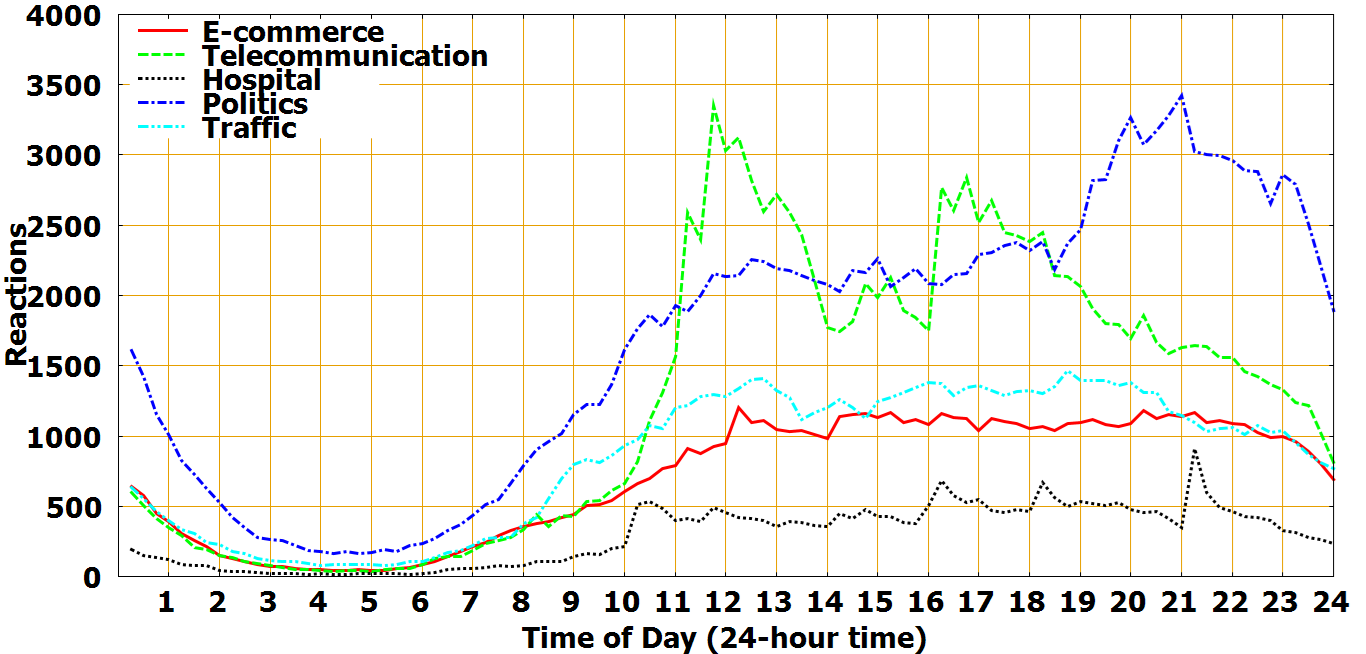}
  \caption{Audience Reaction Pattern on Daily basis}
  \label{daily}
\end{figure}

For daily analysis, we analyze the reaction behavior for 
all the above mentioned five categories, for 24 hours period over a duration of 5 years.  
Unlike Figure~\ref{fig:arr} which shows audience reaction behavior of individual pages, 
Figure~\ref{daily} shows the aggregated audience reaction behavior of the categories.

We observe in Figure~\ref{daily} that categories can have audience reaction
in different ways, such as multiple peaks, single peak and uniform peak during
a day. 

First, we analyze the categories which have multiple reaction peaks in a day (i.e., traffic, telecommunication).  
Reactions on traffic category are high 
during the start of office hours (11 AM) and end of office hours (6 PM to 8 PM). 
One of the reasons is that there is high traffic in these time-periods 
and people react in Facebook pages about the traffic problems which they have faced 
while going or coming back from offices. Similarly, telecommunications category has two peaks in the day: first is around 10 AM to 12 AM and second is around 4 PM to 6 PM. 
 One of the reasons for this is that most of the people interact to social media pages in the morning to complain about an issue or to get the information related to tariffs, vouchers, special offers so that they can fill their balance and can use it throughout the day without out of balance problem. Some people prefer to do the same activity in the evening so that they can talk to family, friends, and relatives in the night when they become free from regular activities. 

E-commerce category has uniform reactions from 12 PM to 10 PM (mostly during office hours) and drops after these hours.
One of the possible reason is that people usually take the opinion of their colleagues and friends working in the same office about the product. If they found any issue, they often bring it to the notice of that e-commerce business immediately using Facebook page due to its quick response.  

Pages related to politics and hospitals have single reaction peak per day. 
There is a high peak of audience reactions on politics category between 8 PM to 9 PM. 
One of the possible reasons is that people become free from their daily work by this time and 
spend some time in knowing the political updates which are posted during the daytime. 
Similarly, people complain more about hospital related issues in the evening which they faced during the daytime.

\subsubsection{Weekly Analysis}
In weekly analysis, we analyze audience reaction behavior on 
two categories namely telecommunication and traffic over the period of a week. 

\begin{figure}[h]
  \centering
  \includegraphics[width=9cm, height=5cm]{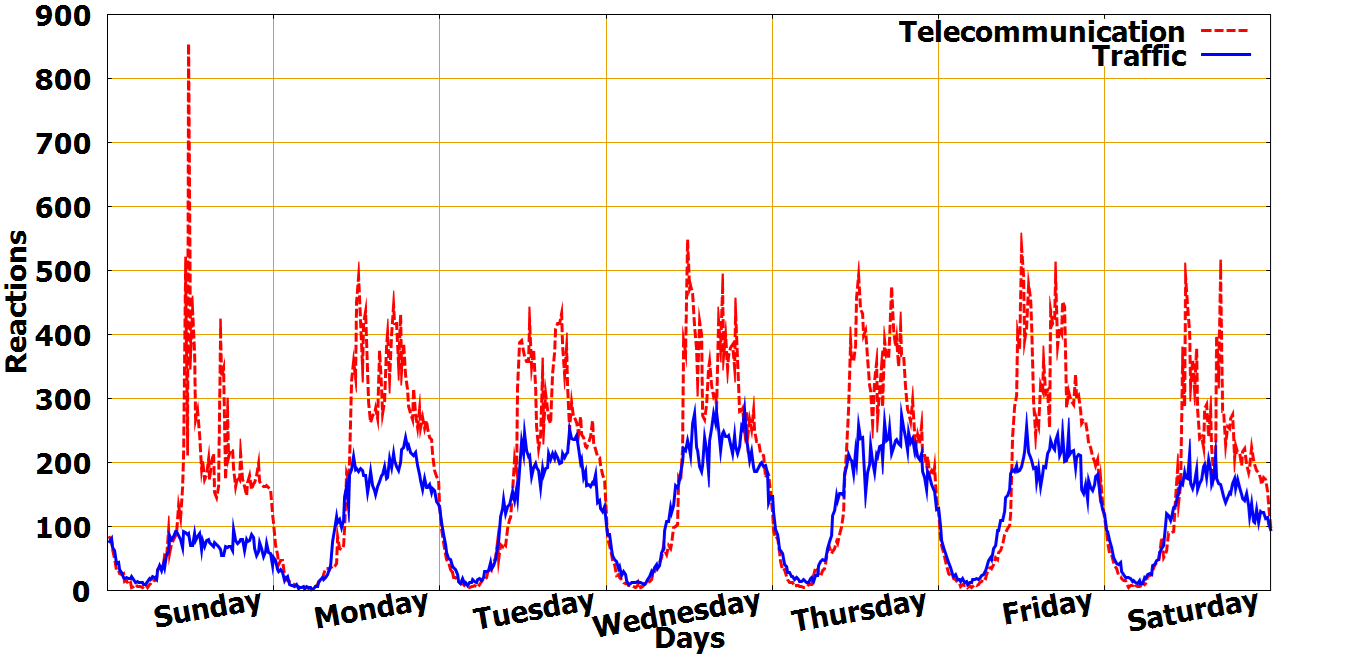}
  \caption{Audience Reaction Pattern on Weekly basis}
  \label{weekly}
\end{figure}

Telecommunication category has the highest peak during Sundays compared to other days of the week.
One of the reasons is that most of the people are free on Sundays 
and they prefer to fill their mobile and data balances. People react more to posts related to telecommunication such as special offers, vouchers during these days. 
Therefore, it is better to post important updates and offers on Sundays 
instead of other weekdays to get a large number of audience reactions.

Reactions on traffic category are high 
during working days and drop slightly during weekends. 
One of the reasons is that people 
do not go to offices on weekends as they have holidays. 
Audience reactions drop to half during Sundays compared to other days of the week 
because even on Saturday some people 
still go to offices, but most of the people don't go to offices on Sunday. Most of the people stay at home and react less in traffic pages during weekends.

\subsubsection{Monthly Analysis}
In monthly analysis, we present audience reaction pattern 
on two categories namely e-commerce and politics over the period of a year.

\begin{figure}[h]
    \centering
  \includegraphics[width=9cm, height=5cm]{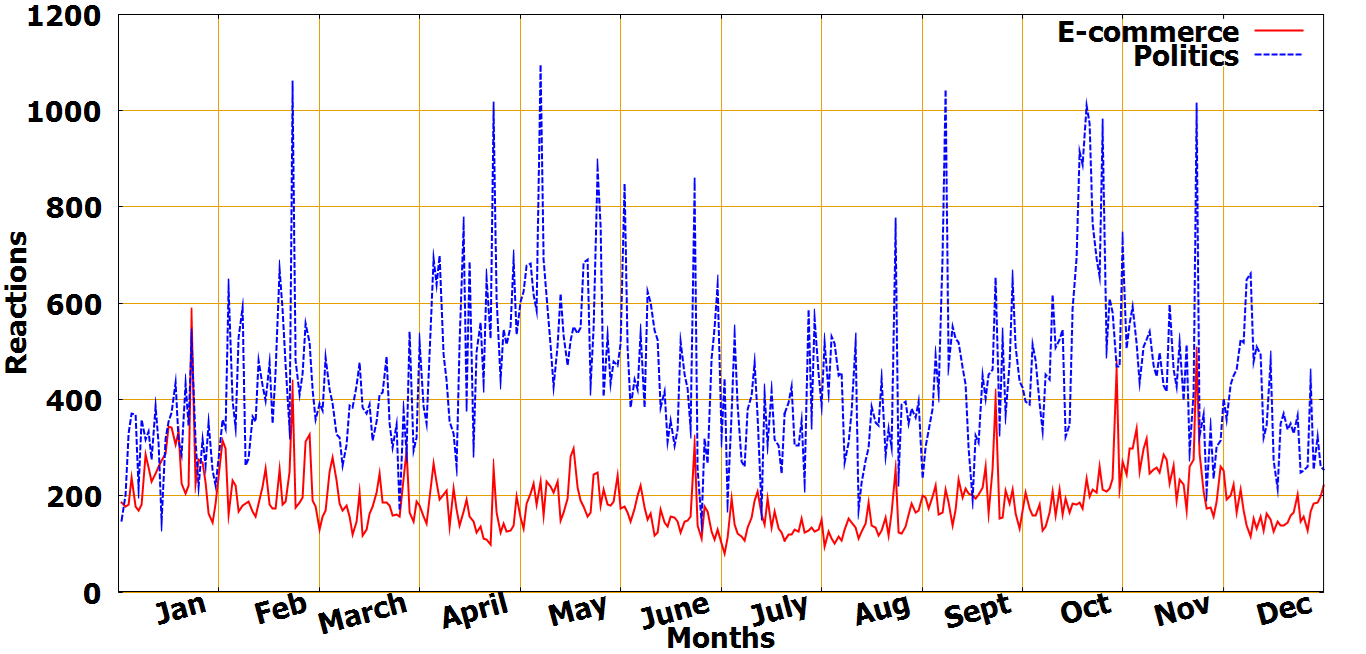}
    \caption{Audience Reaction Pattern on Monthly basis}
    \label{monthly}
\end{figure}

As can be seen in Figure~\ref{monthly}, politics pages received more number of reactions in the months of April and May. 
One of the possible reasons is that the politics pages 
included in dataset had their elections in these months. People are more active on social media pages during election period. 
The peak in month of October and November is due to the introduction
of new fiscal policies. People react more about the advantages and disadvantages of new policies through the social media pages during these periods.  

E-commerce category has more number of reactions during the month of October and November 
because these are the festive months in India and people buy 
new goods on the occasion of festivals. 
Increase in reactions during mid of December, January is due to
Christmas and end of the year sale. 
These are the festive occasions and people like to purchase new items during these occasions. 
They would be interested to know about offers and sales during these periods. 
If an e-commerce page post a news related to these sales and offers, people tend to react on it. 
Moreover, during these sales, lots of people purchase new items and a large fraction of these people face the problems such as delivery issue, product issue, etc. People share their experiences\footnote{Today discounts looks impressive hope big billion days rock coming days?} and complaints\footnote{I ordered product exchange offer honor b 15oct 2015 but yesterday cancelled order without information.} about the issue through the social media pages of e-commerce.

\subsection{Audience Engagement with Contents}
\label{rtp:sec_aewc}
In this section, we show through empirical results that audience engagement depends on the type of content. Facebook page admins
create different types of content, such as photo, video, link, and
status. Some of these types of content receive more number of audience reactions compared to others. Pages can achieve higher audience engagement by creating contents of the type that receives more audience reaction. The results in Table~\ref{table:4} are based on the dataset  mentioned in Section~\ref{sec:rtp_dataset}.

\begin{table}[h]
\begin{center}
\begin{tabular}{|c|l|l|l|}
\hline
\hline
\thead{\bf Content type} & \thead{\bf Posts \%} & \thead{\bf Reactions \%} \\
 \hline\hline
 Link & 78.60\% & 54.16\% \\ 
 \hline
 Photo & 8.55\% & 18.46\% \\
 \hline
 Status & 1.59\% & 1.21\% \\
 \hline
 Video & 11.26\% & 26.17\% \\ [1ex] 
 \hline
\end{tabular}
\caption{Posts and reactions of different types of contents} 
\label{table:4}
\end{center}
\end{table}

In Table~\ref{table:4}, for each content type, the second column shows the percentage of posts created by all the pages of that type, and the third column shows the percentage of reactions received by all the posts of that type. From the first row, we observe that although pages post 78.60\% of the content as links, they get only 54.16\% reactions from such content. In other words, links give less reaction (or audience engagement) per post. On the other hand, pages post only 8.55\% and 11.26\% content as images and videos, which brings 18.46\% and 26.17\% audience reaction respectively. Videos can bring highest audience engagement.

\section{Related Work}
\label{sec:rtp_rw}
One can use influential users to increase content visibility. There are
existing works~\cite{bapna2015your,bharathi2007competitive,chen2009efficient,kao2016mining,kempe2003maximizing,probst2013will} 
that find influential users in the social network, and use
these influential users to spread information. 
Recently, researchers~\cite{rao2015klout} at Klout developed a influence scoring system that measures influence of users for targeted search and marketing. 
In this paper, we look at the problem of
spreading information by finding what is the right time to create a post so that
the post can get high audience reactions. If a post is getting high audience reactions, it will
automatically become popular because it would be shown at the top of audience news feed.  
Our approach is complementary to the existing approach of using influential users to increase content visibility.

To spread information in a social network, we need to understand the flow and 
diffusion of information in the social network~\cite{centola2010spread,cheng2014can,farajtabar2016multistage,guille2013information,lerman2010information,leskovec2007patterns}. It requires an understanding of the topological structure and temporal 
characteristics of the
social network~\cite{kwon2013prominent,leskovec2009meme,pei2013spreading,szabo2010predicting,tsytsarau2014dynamics}. For example, if a user connected to many users, 
posts some information it will automatically reach to many users. In this paper, 
we do not use topological structure. However, we use the right time to 
post to increase information diffusion. In future work, we would look
at combining topological analysis with our approach to get even
higher information diffusion.

To understand audience reaction behavior, we need to understand the user dynamics
in social networks~\cite{das2014modeling,kwak2010twitter,lehmann2012dynamical,wu2013arrival,yu2017temporally}. User dynamics changes
across different social networks~\cite{li2014social}. For example, in Twitter, the lifetime of content is quite short compared to other social networks. Some of the
topics end in just 20-40 minutes~\cite{asur2011trends}. 
Wu et al.~\cite{wu2011says} show that regardless of the type of content, 
all contents have a very short life span that usually drops exponentially after a day. One of the most recent work in user dynamics carried out by Rizoiu et al.~\cite{rizoiu2017online} models the popularity dynamics of online items. They investigate the factors that  influence the forecast of future popularity under promotion and use it to quatify expected attention generated by external promotion. 
In this paper, we study the user dynamics in Facebook pages. This social network
is somewhat different compared to Twitter. Here user dynamics is
somewhat slower compared to Twitter. 
Moreover, there have been few studies on finding the right posting schedule for social network users which stated that posting time also depends on the user dynamics~\cite{karimi2016smart,spasojevic2015post,zarezaderedqueen}. 
However, these works mainly focused on finding the right posting schedule for individual
users in social network. Their posting schedules are derived based and the users’ social
connections and locations. They do not look at many other features that can affect
audience reactions, such as features about the content~\cite{biswas1994mood,esiyok2014users,reis2015breaking} or features about the content creator~\cite{yan2012better}. Our work is complementary to existing approaches that attempts
to find the right time to spread the information of social media brand pages towards a
large audience.

In this paper, we look 
at Facebook pages, which has follower-following type of relationship. A page
can have unlimited number of followers, whereas a user can have at most
few thousand friends~\cite{taylor2011friends}. We look at large number of features 
to find the best posting schedule. 
In addition to compute schedule for individual
pages, we also look at the problem of finding schedule for a group
of pages with similar audience reaction.

\section{Conclusion}
\label{sec:rtp_cnf}
In this paper, we looked at the problem of how to increase
the visibility of a content in social media brand pages by posting messages at a time
that increases the likelihood of getting audience reactions.  
We analyzed user dynamics for individual Facebook pages as well as 
for a group of Facebook pages, with similar reaction profile, which we call page category. 
We proposed six schedules for getting high audience reaction, 
amongst which the best schedule leads to seven times higher reaction gain. 
We presented interesting audience reaction patterns in the 
form of daily, weekly and monthly temporal patterns.

\balance
\bibliographystyle{splncs03}
\bibliography{bibliography} 
\end{document}